\documentclass[letter]{aa} 

\usepackage{graphicx}
\usepackage{txfonts}
\usepackage{color}
\definecolor{applegreen}{rgb}{0.55, 0.71, 0.0}
\usepackage{ulem}
\usepackage{amssymb}
\usepackage{xspace}

\newcommand{\inte}{{INTEGRAL}\xspace}

\newcommand{\maxij}{MAXI J1348$-$630\xspace}
\newcommand{\swiftj}{\textit{Swift} J1727.8$-$1613\xspace}

\usepackage{subcaption}
\begin{document} 

 \title{INTEGRAL/IBIS polarization detection in the hard and soft intermediate states of \textit{Swift} J1727.8$-$1613}
 
\titlerunning{Polarized hard tails in  Swift J1727.8$-$1613}

   \author{T. Bouchet\inst{1},  J. Rodriguez\inst{1}, F. Cangemi\inst{2}, P. Thalhammer\inst{3}, P. Laurent\inst{1},  V. Grinberg\inst{4}, J. Wilms\inst{3} \& K. Pottschmidt\inst{5,6}
   }
\authorrunning{Bouchet et al.}
 \institute{Universit\'e Paris Cit\'e, Universit\'e Paris-Saclay, CEA, CNRS, AIM, F-91191 Gif-sur-Yvette, France 
            \and APC, Universit\'e Paris Cit\'e, CNRS, CEA, Rue Alice Domont \& L\'eonie Duquet, 75013 Paris, France 
            \and Dr.\ Karl Remeis-Observatory and Erlangen Centre for Astroparticle Physics, Friedrich-Alexander-Universit\"at Erlangen-N\"urnberg, Sternwartstr.~7, 96049 Bamberg, Germany
            \and European Space Agency (ESA), European Space Research and Technology Centre (ESTEC), Keplerlaan 1, 2201 AZ Noordwijk, The Netherlands
            \and NASA Goddard Space Flight Center, Astrophysics Science Division, 8800 Greenbelt Road, Greenbelt, MD 20771, USA
\and CRESST and Center for Space Sciences and Technology, University of Maryland Baltimore County, 1000 Hilltop Circle, Baltimore, MD 21250, USA}

   \date{}

 
  \abstract
{}
{Soft $\gamma$-ray emission  (100 keV -- 10 MeV) has previously been detected in the hard state of several microquasars. In some sources, this emission was found to be highly polarized and was suggested to be emitted at the base of the jet. Until now, no $\gamma$-ray polarization had been found in any other state.}
{Using \inte/IBIS, we studied the soft $\gamma$-ray spectral and polarization properties of \swiftj throughout its outburst.}
{We detect a highly polarized spectral component in both the hard intermediate state and the early stages of the soft intermediate state above 210\,keV. In the hard intermediate state, the polarization angle significantly deviates from the compact jet angle projected onto the sky, whereas in the soft intermediate they are closely aligned. This constitutes the first detection of jet-aligned polarization in the soft $\gamma$-ray for a microquasar. We attribute this polarized spectral component to synchrotron emission from the jet, which indicates that some of the jet might persist into the softer states.}
{}

\maketitle
%

\section{Introduction}
Black hole X-ray binaries (BHXBs) are compact binary systems that host a (candidate) black hole accreting matter from a companion star. They are commonly found to be sources of powerful jets and are therefore dubbed microquasars. Most BHXBs are transient sources seen to transit between various spectral states. In the (low) hard state (LHS), the 1--100 keV spectrum is phenomenologically described  by a power-law with an exponential cutoff, interpreted as the thermal Comptonization of photons from an accretion disk by hot electrons. In the (high) soft state (HSS), the accretion disk emission becomes dominant below 10\,keV and a weak power-law component may be present above 10\,keV. Between these two states are the hard and soft intermediate states \citep[respectively, the HIMS and SIMS; see, e.g.,][]{Belloni_2010, Remillard_2006}. 

In several BHXBs, an additional component is sometimes seen above $\sim$ 200\,keV \citep[e.g.,][]{grove98, Jourdain_2012}. It is usually brighter in the LHS and the HIMS \citep[e.g.,][and references therein]{Rodriguez_2015,Cangemi_2023a}. This ``hard tail'' could originate from synchrotron emission at the base of the jet \citep{Markoff_2005} or from a hybrid corona of thermalized and non-thermalized electrons \citep{Gierlinksi_1999}. 
Polarization above  ${\sim} 400$\,keV was first detected in the persistent microquasar Cyg X-1 by the Imager onBoard the \inte Satellite \citep[IBIS;][]{Laurent_2011} and \inte Spectrometer \citep[SPI;][]{Jourdain_2012}. The source was later confirmed to be polarized in the LHS \citep{Rodriguez_2015}, and in the HIMS with {AstroSat}/CZTI \citep[][]{Chattopadhyay_2024}. These three independent instruments gave compatible polarization angles (PAs) and polarization fractions (PFs) in their overlapping energy ranges
(see Fig. 3 of \citealt{Chattopadhyay_2024}).
LHS polarization was also found in the transients \maxij and MAXI J1820+070 \citep[][hereafter CRB23]{Cangemi_2023a}.

\swiftj (J1727 hereafter) was discovered as a $\gamma$-ray burst by \textit{Swift} in August 2023 \citep{page_2023gcn} and identified as a BHXB with follow-up observations from the Monitor of All-sky X-ray Image (MAXI) \citep{s1727_maxi_atel}. The source was extremely bright at peak (up to 8\,Crab in the 2--20\,keV band), making it one of the brightest BHXBs ever detected. The source was intensely monitored from the radio to $\gamma$-rays and notably showed the presence of a hard tail in the LHS and HIMS \citep{Cangemi_2023b}. The compact jet has been resolved in the radio, and its position angle was estimated to be $ - 0.60 \pm 0.07^\circ$ \citep[][]{Wood_2024}.

This Letter is organized as follows: We describe the \inte observations and data reduction in Sect.~\ref{sec:reduction}, focusing on the calibration necessary for the most recent data. We then present the results of our phenomenological spectral study and polarization analysis in Sect.~\ref{sec:results} and discuss our findings in Sect.~\ref{sec:discussion}.

\section{Data reduction and analysis} \label{sec:reduction}
The journal of the \inte\  observations of J1727 is reported in Table \ref{tab:observations}. 
We used data from IBIS \citep{Ubertini_2003} and from the Joint European Monitor for X-rays \citep[JEM-X;][]{Lund2003}. 
IBIS consists of two detector planes, the IBIS Soft Gamma Ray Imager \citep[ISGRI;][]{Lebrun_2003}, 
sensitive in the 30--500\,keV range, and the Pixellated Imaging Caesium Iodide Telescope \citep[PICsIT;][]{labanti03}, sensitive in the 150\,keV--10\,MeV range. These detectors can either be used independently or together as a ``Compton telescope,'' which allows the 200\,keV--10\,MeV polarization to be probed (Sect.~\ref{sec:compton}).

\subsection{JEM-X and ISGRI data reduction}
ISGRI products were extracted with the \inte Off-line Scientific Analysis (OSA) software  v11.2, following standard reduction processes\footnote{\url{https://www.isdc.unige.ch/integral/download/osa/doc/11.2/osa_um_ibis/man_html.html}}. For observations after 2020, the ISGRI calibration is (currently) not up to date. We thus created home-corrected ancillary response files using recent ISGRI spectra of the Crab that were adjusted to match the corresponding SPI spectra. The correction is particularly important in the 30--100\,keV band, where ISGRI suffers most from CdTe aging. Fluxes were obtained for the 30--50, 50--100, and 100--200\,keV ranges from individual pointings, the $\sim$1700\,s long ``Science Windows'' (ScWs). We first stacked all ScWs from the same satellite revolution to obtain source spectra. We added a systematic error of 3\% to each spectral channel as recommended in the User Manual.

To better constrain the continuum, we added the 5--20\,keV spectra from JEM-X unit~1, applying the same revolution-based stacking as for ISGRI. We included 3\% systematic errors on the spectra that we obtained with the procedure described by CRB23\footnote{\url{https://www.isdc.unige.ch/integral/download/osa/doc/11.1/osa_um_jemx/man.html}}.

\subsection{Compton mode and polarization} \label{sec:compton}
Photons with energy $\gtrsim$ 200\,keV can be Compton-scattered by ISGRI and absorbed by PICsIT. The energy of the incident radiation can then be deduced from the energy deposited in both layers of IBIS \citep[see][]{Forot_2007, Forot_2008}. It is also possible to reconstruct the azimuthal scattering angle, $\phi$, from the interaction points on both detectors, such that the PA of the incident radiation can be reconstructed.

Polarization measurements of the Crab between 250 and 450\,keV are found to be compatible with independent SPI measurements when applying these corrections \citep{Bouchet_2024}. This allowed us to explore the polarization in the 210--300\,keV range with \inte(/IBIS) for the first time.  Appendix~\ref{sec:A-Polar} gives a brief overview of the principles of polarization measurements and the estimates of significances and upper limits.

\section{Results}\label{sec:results}
\subsection{State classification and spectral evolution}\label{sec:spectra_isgri_jemx}

\begin{figure*}
    \centering
    \includegraphics[width=\linewidth]{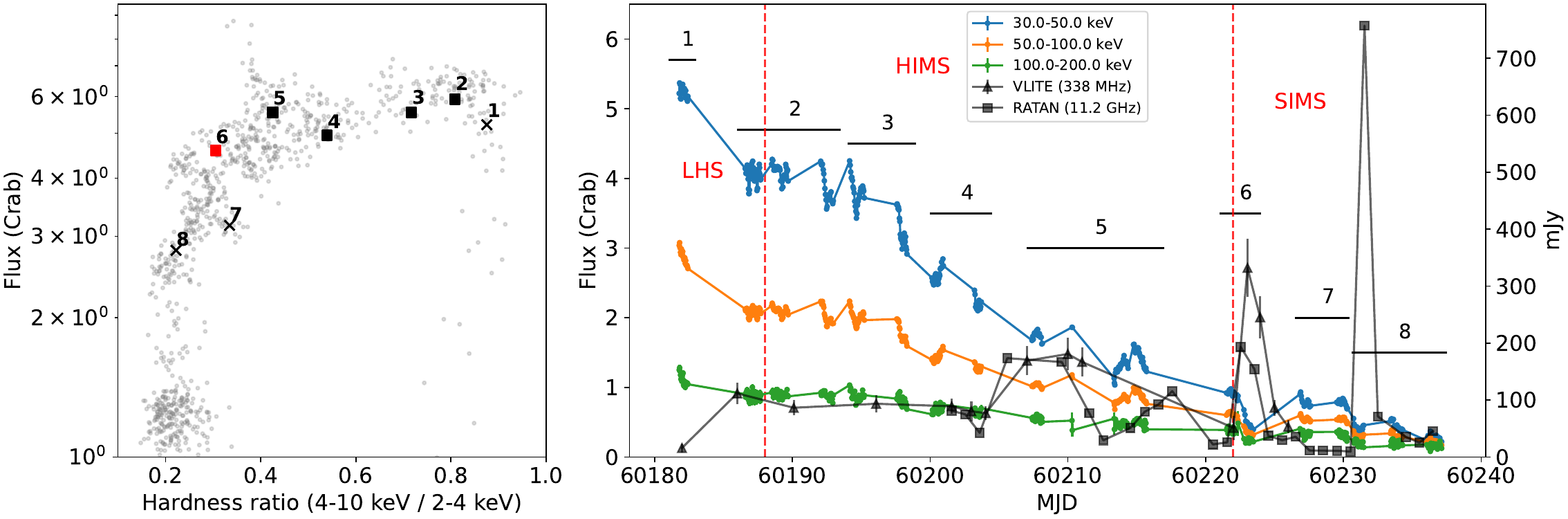}    
    \caption{Evolution of J1727 outburst. Left: Hardness intensity diagrams from the MAXI/GSC data. Data points closest in time to \inte observations are highlighted as crosses for those with no polarization, and as black (respectively\ red) squares for those with a PA misaligned (respectively aligned)  with the jet angle. \textbf{Right:} \inte/ISGRI ScW-based and radio (VLITE and RATAN) light curves of J1727. In both plots the numbers show the locations of our grouped observations.}
    \label{fig:S1727_lc_states}
\end{figure*}

Figure~\ref{fig:S1727_lc_states} shows the light curves of 
the entire outburst of J1727 with data from IBIS/ISGRI (20--200\,keV), the 338\,MHz VLA Low-band Ionosphere and Transient Experiment \citep[VLITE;][]{VLITE_ATel}, and the RATAN-600 radio telescope at 11.2\,GHz \citep[extracted from][]{Ingram_2023}, as well as the corresponding hardness intensity diagram obtained with the MAXI data.
We used the spectral state classification and dates of transitions reported from the analysis 
of the Neutron Star Interior Composition ExploreR (NICER) data \citep{nicer_qpo_hard, nicer_qpo_im, nicer_qpo_soft}. The source 
first rises in the LHS (revs.\ 2678 and 2680; see Table~\ref{tab:observations}) and 
starts transiting to the HIMS around MJD\,60185 (rev.\ 2681). The \inte\ light curves 
decrease rather smoothly until $\mathrm{MJD}\sim 60221$ (rev.\ 2694), at which point we see the occurrence of $\sim$150--200\,mJy 11.5\,GHz radio flares (MJDs $\sim$60205 and 60221), which occurred around the time of the transition to the SIMS. The source had small re-brightenings at high energies around MJDs 60226--60229 (revs.\ 2695 and 2696)
before experiencing a bright radio flare (MJD\,60230) that preceded its global decrease.  

\begin{figure*}
    \centering\includegraphics[width=19cm]{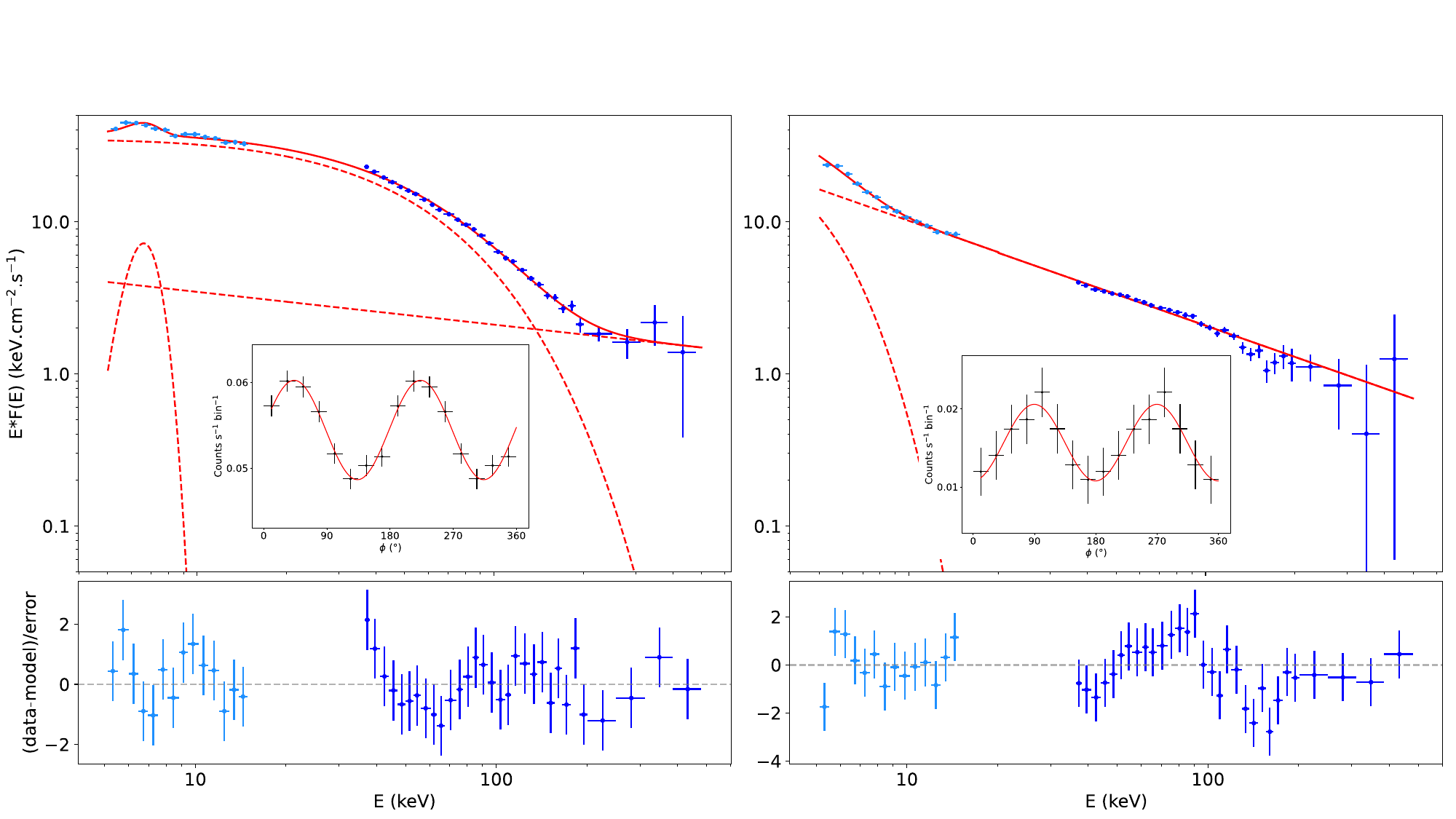}
\caption{JEM-X (light blue) and ISGRI (dark blue) spectra. The continuous red lines show the total fit and the dashed red lines the various spectral components. The inserts show the 210--250\,keV polarigrams, with the data points in black and the model in red. \textbf{Left:} Obs.~3 (HIMS).
    \textbf{Right:} Obs.~6 (SIMS).}
    \label{fig:spec-pola}
\end{figure*}

We performed a phenomenological analysis of the 5--15\,keV and 35--500\,keV JEM-X unit 1 and ISGRI spectra on a revolution-by-revolution basis with simple models of the XSpec software \citep{Arnaud_1996}: a power-law (\textsc{powerlaw}, hereafter PL) and a cutoff power-law  (\textsc{cutoffpl}, hereafter CPL) for the continuum. A normalization constant (frozen to 1 for IBIS) was included to account, in particular, for the differences in IBIS and JEM-X exposure times, as well as for imperfections in source reconstruction due to its large brightness. The best model was retained on the basis of the $\chi^2$ value of the fit, and, when similarly good fits were obtained with various combinations, we retained the one with the JEM-X normalization constant closest to 1. We then combined spectra that were close in time and spectral behavior to obtain a better statistical significance. Some fits have larger $\chi^2$ values, which can be due to the source becoming fainter at the end of the outburst and thus the relative background increasing (due to the other sources in this crowded field), which may produce the residuals at higher energies. The $\chi^2$ values are nevertheless acceptable for our phenomenological analysis. An overview of the state-based stacked observations is given in Table~\ref{tab:observations}.
Observations~1-5 and 7 were fitted with a CPL, covering the spectra up to $\sim$ 100 keV, and an additional PL for Obs.~1-4, covering the excess above $\sim$ 100 keV. Observations~6 and 8 are best described with a single PL. A Gaussian line (\textsc{gauss}) representing the Fe K$\alpha$ line was added for Obs.~2-5 (see Fig.~\ref{fig:spec-pola}, left), and an accretion disk (\textsc{diskbb}) fixed at an inner disk temperature of 1\,keV was added for Obs.~6-8 (see Fig.~\ref{fig:spec-pola}, right). The results of these phenomenological fits and their statistical significance are reported in Table~\ref{tab:spectral}.

   \begin{table*}
    \caption{Parameters from 5--500 keV spectral fits for the observation groups defined in Fig.~\ref{fig:S1727_lc_states}. Missing entries indicate that a component was not required. Errors are the 90\% confidence intervals.}
    \label{tab:spectral}    \centering
    {\renewcommand{\arraystretch}{1.3} 
    \begin{tabular}{ccccccccc} 
            Group \# & C$_{JEM-X}$ & $\Gamma_\mathrm{CPL}$ & E$_\mathrm{cut}$ & Flux$_{\mathrm{CPL},30-300\,\mathrm{keV}}$ & $\Gamma_\mathrm{PL}$ & Flux$_{\mathrm{PL},30-300\,\mathrm{keV}}$ & $\chi_\nu^2 (\mathrm{dof})$ & $\frac{F_{\mathrm{PL},210-300}}{F_{\mathrm{Total},210-300}}$  \\
            & & & [keV] & [$10^{-9}\,\mathrm{erg}\,\mathrm{cm}^{-2}\,\mathrm{s}^{-1}$] &  & [$10^{-9}\,\mathrm{erg}\,\mathrm{cm}^{-2}\,\mathrm{s}^{-1}$] &  &  \\
    \hline

 1  & 1.79 & 1.47$_{-0.13}^{+0.07}$  & 33$\pm2$ & 46  & 2.1$_{-0.3}^{+0.2}$ & 7.1  & 1.48 (39) & 90\%\\
 2$^\dagger$  & 0.72 & 2.00$_{-0.11}^{+0.04}$ & 44$_{-5}^{+4}$ & 33  & 2.1$_{-0.4}^{+0.3}$ & 7.6   & 1.08 (40) & 90\%\\
 3$^\dagger$  &  0.94 & 2.24$_{-0.13}^{+0.07}$  & 41$_{-4}^{+4}$ & 28 & 2.2$_{-0.3}^{+0.2}$ & 8.9& 0.84 (40)  & 91\% \\
 4$^\dagger$  & 0.89 & 2.11$\pm$0.03 & 57$_{-11}^{+7}$ & 23 &  1.4$_{-0.6}^{+0.8}$ & 3.5 & 1.03(40) & 81\% \\
 5$^\ast$ & 1.02 & 2.46$\pm$0.07 & 152$_{-20}^{+26}$ & 17  & -  &  - & 1.22 (36)& - \\
 6 $^\ddagger$  & 0.91 & - & - & - & 2.68$\pm$0.02 & 9.6   & 1.24 (41)& - \\
 7 $^\ddagger$  & 1.05 & 2.51$\pm$0.10 & 274$_{-76}^{+163}$ & 10 &  - & - & 1.31 (40) & -  \\
 8 $^\ddagger$  & 0.91 & - & -& - &  2.64$\pm$0.03 & 5.6 & 1.77 (41) & -  \\
\hline
    \end{tabular}}
\tablefoot{$^\ast$JEM-X spectrum ignored below 7\,keV due to large instrumental residuals. $\dagger$ Gaussian emission line with fixed center at 6.4\,keV. $^\ddagger$Thermal disk with a fixed temperature of 1\,keV.}
\end{table*}

\subsection{Polarization properties}
\begin{figure}
    \centering
    \includegraphics[width=\linewidth]{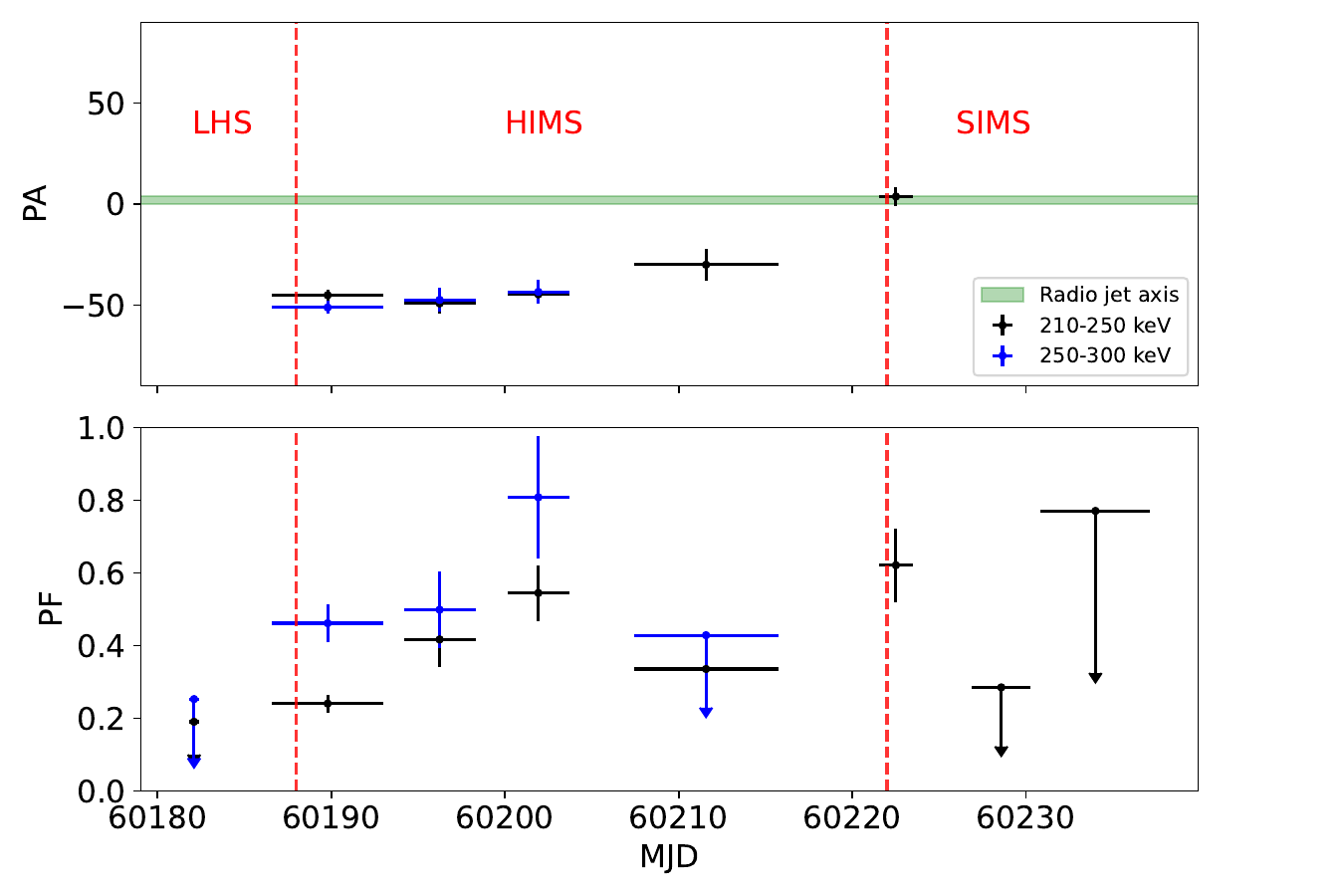}
    \caption{Temporal evolution of the J1727 PA (upper panel) and PF (lower panel) in the 210--250\,keV (black) and 250--300\,keV (blue) bands. The radio jet PA is shown in green. The upper limits of the 250--300\,keV PF of Obs.~6 to~8 are close to 1 and not shown.}
    \label{fig:S1727_pola_evolution}
\end{figure}

Figure \ref{fig:S1727_pola_evolution} shows the temporal evolution of the PF and PA of our grouped observations in the 210--250\,keV and 250--300\,keV ranges. All polarimetric values along with their S/Ns can be found in Table \ref{tab:polar}. All errors are given for a 68\% confidence interval.
No polarization is detected in Obs.~1. The exposure time is shorter than for the next combined observation, but the much higher flux still allows us to put satisfying upper limits of 19\% (respectively 25\%) in the 210--250\,keV bands (respectively\ 250--300\,keV)  for this LHS observation. We detect polarized signals in all HIMS observations with PF$_{250-300}\geq$PF$_{210-250}$ and a rather stable and similar PA in the two energy ranges in Obs.~2 to~4 (around $-45^\circ$). The 210--250\,keV (respectively\ 250--300\,keV) PF increases up to $54 \pm 8$\% (respectively\ $81 \pm 17$\%) in Obs.~4. Significant evolution is visible in Obs.~5,  where $\mathrm{PA}_{210-250} = -30 \pm 7^\circ$ with a decrease in PF. A more detailed view of the energy dependence is shown in Appendix~\ref{sec:C-PF}.

In Obs.~6 (transition to the SIMS), we detect a strong polarized signal in the 210--250\,keV band ($p_\mathrm{unpolarized} = 0.051\%$) with $\mathrm{PF}_{210-250}=62 \pm 10\%$ and $\mathrm{PA}=3.8 \pm 4.6^\circ$.
We do not detect polarization in Obs.~7  with $\mathrm{PF}_{210-250}\leq29\%$. In Obs.~8 we find a quite unconstraining $\mathrm{PF}_{210-250}\leq$77\%, but the source significance is also much lower.

\section{Discussion}
\label{sec:discussion}

\subsection{Hard-tail origin}

Our spectral analysis of J1727 confirms the presence of a $\gtrsim$100\,keV hard tail \citep{Cangemi_2023b} persisting throughout the whole HIMS (Table~\ref{tab:spectral}) in addition to a CPL, interpreted as thermal Comptonization, dominating below 100\,keV. At the transition to the SIMS (Obs.~6), the CPL is replaced by a PL extending out to about 400\,keV (Fig.~\ref{fig:spec-pola}, left). Observation~7 can be fitted with a CPL, but the cutoff energy is above $\sim$250\,keV and not well constrained, meaning it could be modeled by a simple PL like Obs.~6 and 8 if we had access to better data at higher energies.
The presence of a hard tail on top of Comptonization and the persistence of a $\gtrsim$100\,keV PL emission during softer states is similar to the behavior of Cyg X-1 \citep{McConnell_2002}, MAXI J1348$-$630, and MAXI J1535$-$571 (CRB23). The CPL components seems to disappear (quasi-)simultaneously with a radio flare (Fig.~\ref{fig:S1727_lc_states}), which might  indicate that (part of) the Comptonizing plasma is ejected \citep[see, e.g.,][for similar behavior in XTE J1550$-$564 and GRS 1915+105]{rodriguez_2003,rodriguez_2008}. 

The polarization properties of J1727 are different in both states. The PA in the HIMS is stable with time and energy (Figs. \ref{fig:S1727_pola_evolution} and \ref{fig:S1727_him_pola_spec}), whereas the PF increases with energy and varies with time. 
The polarization properties change in Obs.~5, with a significant evolution of the PA and a decrease in the PF. While the misalignment of the PA with the radio jet may call into question the association of the hard tail with the radio jet, similar discrepancies are also seen in, for example, Cyg X-1 \citep{Laurent_2011}. The level of the PF in all HIMS observations is compatible with 
the maximum PF expected from synchrotron radiation in an ideal case \citep[$\sim$75\% for a $\Gamma\sim2.1$ PL;][]{synchrotron_pf}, and 
it is challenging to reconcile PF values above 40\% with another emission mechanism. 

The difference in position angle between the jet and polarization in the HIMS was 
discussed by \cite{Russell_2013}, who suggest two main possibilities: a misaligned limb-brightened spot on 
the ridge of the jet combined with a large ($> 30^\circ$) opening angle at its base (see Fig.~3 of their article), or an electron distribution with a sharp spectral change and a nonuniform pitch-angle distribution within the jet, as shown by \cite{Bjornsson_1985} for blazars.

The detection of a polarized signal near the probable transition toward the SIMS with a large PF, 210--250\,keV  (the signal is too weak at higher energies for a meaningful measure), is new and unexpected. In this band, the PA is aligned with the compact jet axis resolved in the radio. The quasi-simultaneous occurrence of a radio flare (${\sim} 200$\,mJy at 11.2\,GHz), probably associated with a discrete ejection of material, suggests an 
association of the polarized high-energy PL component with the jet. The 62\% PF we measure is also compatible with the maximum expected polarization (80\%) from synchrotron emission with a $\Gamma=2.7$ PL. This favors the jet as the source of the SIMS PL emission, as proposed by \citet{Jourdain_2014} for Cyg X-1. We also note that for synchrotron emission, the magnetic field interacting with the electrons has to be distributed orthogonally to the jet axis, and the inclination of the jet (i.e., the angle between the jet axis and the line of sight) will likely affect the PF. According to \citet{Laing_1980}, an inclination angle of at least ${\sim} 60^{\circ}$ is required to achieve this level of polarization for a randomly distributed magnetic field in a slab orthogonal to the jet. Blazars dominated by synchrotron emission in the X-ray were observed to be polarized at a few percent \citep{Middei_2023} and up to 30\% \citep{Kouch_2024} despite their intrinsically low inclination.

\subsection{Simple phenomenological model}

To explain the evolution of polarization, we used a phenomenological two-population toy model on top of the usual inverse Compton emission. This approach is compatible with both hypotheses of \cite{Russell_2013} and aims to provide a simplified framework for our discussion:
population~1 interacts with a perpendicular magnetic field, producing synchrotron emission with $\mathrm{PA}_1$ parallel to the jet; and  population~2 produces synchrotron emission polarized at a shifted angle $\mathrm{PA}_2$ compared to population~1 because of a different electron pitch angle. This simplistic representation is also motivated by the persistence of radio emission throughout the whole outburst (Fig.~\ref{fig:S1727_lc_states}), which may indicate the persistence of some kind of jet.

We can compute different polarized emissions with different PAs, PFs (assumed constant with energy for simplicity), and energy flux distributions ($f_i = F_i / F_\mathrm{tot}$) added together. From the angle distribution of the emitted photons (Eq.~\ref{eq:1}), we deduce the resulting polarization parameters as a function of energy,
\begin{equation}\label{eq:PA}
\mathrm{PA}(E)=\frac{1}{2}\ \arctan\left(\frac{\sum_{i}f_i(E)\ \mathrm{PF}_i\ \sin{\left(2\mathrm{PA}_i\right)}}{\sum_{i}f_i(E)\ \mathrm{PF}_i\ \cos{\left(2\mathrm{PA}_i\right)}}\right)
\end{equation}
and
\begin{equation}\label{eq:PF}
\mathrm{PF}(E)=\frac{\sum_{i}f_i(E)\ \mathrm{PF}_i \cos{(2\mathrm{PA}_i)}}{\cos{(2\mathrm{PA}(E))}}
.\end{equation}

During Obs.~1 (LHS with PF< 25\%), the radio flux at 338 MHz is lower by a factor of at least 5 compared to the next few observations in the HIMS  (see Fig. \ref{fig:S1727_lc_states}, right). The hard X-ray flux, on the other hand, is much higher, especially below 100\,keV. We can interpret this as the weakly polarized Comptonized medium being dominant up to 300 keV in the LHS ($f_\mathrm{comptonized} \sim 1$, PF$_\mathrm{comptonized} \leq 20 \%$), at which time  the polarized synchrotron emission from the jet is low. \citet{Chattopadhyay_2024} reached the same conclusion regarding the pure hard state of Cyg X-1 ($\Gamma<1.78$), for which they find a PF upper limit of 7\% in the 100 -- 380\,keV energy range. Unfortunately, our spectral fit has a poor normalization constant between JEM-X and ISGRI (1.79) as well as a large $\chi_2$ value, and the cutoff energy of the CPL is very close to the lower threshold of ISGRI (35\,keV), which prevents us from reliably interpreting the spectrum.

In the HIMS of Obs.~2-4, population~1 is suppressed in the soft $\gamma$-ray ($f_1 \ll f_2$ for $E>210$\,keV) such that $\mathrm{PA}=\mathrm{PA}_2$. In the radio to the IR, where the emission is dominated by synchrotron emission from the jet, $f_1 \gg f_2$, so there must exist an equivalent energy, $E_\mathrm{eq}$, at which $f_1(E_\mathrm{eq}) \sim f_2(E_\mathrm{eq})$ between the IR and soft $\gamma$-ray, but this energy band could be dominated by other emission mechanisms.

In the Obs.~6 SIMS, the second population with misaligned polarization is suppressed ($f_1 \gg f_2$ for $E<E_\mathrm{eq}$) and only the first population emission is seen, therefore producing a parallel angle, $\mathrm{PA}= \mathrm{PA}_1$. This is in agreement with the disappearance of the hard tail seen in the ISGRI spectra. For Obs.~5 at the end of the HIMS, we also find that the 210--250\,keV PA is in between $\mathrm{PA}_1$ and $\mathrm{PA}_2$, meaning that there is possibly $f_1(E_\mathrm{eq}) \sim f_2(E_\mathrm{eq})$ in this energy range, putting the equivalent energy in the 210--250\,keV range.

In soft X-rays, the PA is aligned with the jet with a PF between 3 and 4\% \citep{Veledina_2023, Ingram_2023}, which is interpreted as an effect of the coronal shape. If the highly polarized emission  in Obs.~6 extends to lower energies, it could contribute to the polarization while being smeared out by the accretion disk or corona. Interestingly, optically thin synchrotron emission has already been invoked to explain the radio to X-ray spectra of several HIMS sources \citep[see][for a review]{Russel_2012}.

\section{Conclusion}

For the very first time, a soft $\gamma$-ray polarized signal was found in a BHXB with a PA aligned within a few degrees with the black hole jet axis. This emission coincides with a radio flare and was most likely seen during the transition to the SIMS. The high PF is compatible with optically thin synchrotron emission, and other emission mechanisms should only marginally contribute to the overall flux. In the hypothesis of synchrotron emission, the magnetic field with which the electrons interact should be orthogonal to the jet axis, similarly to the radio synchrotron emission. We also find a misaligned but stable PA in the HIMS, with a PF increasing with energy, hinting at a different electron population related to the hard tail seen above 200\,keV.

We conclude that the medium producing the soft $\gamma$-rays evolves during the outburst, and that the PL observed in the SIMS could originate in part from synchrotron emission, similar to the hard tail seen in the HIMS.


\begin{acknowledgements}
We thank the referee for their careful reading and fruitful comments. We acknowledge partial funding from the French Space Agency (CNES) and from the Deutsches Zentrum f\"ur Luft- und Raumfahrt under contract 50\,OR\,1909. We thank Carlo Ferrigno for his help on the calibration of the IBIS/ISGRI detector. Based on observations with \inte, an ESA project with instruments and science data center funded by ESA member states (especially the PI countries: Denmark, France, Germany, Italy, Switzerland, Spain) and with the participation of Russia and the USA. The material is based upon work supported by NASA under award number 80GSFC21M0002.
\end{acknowledgements}

%
%
\bibliographystyle{aa} 
\bibliography{references}

\begin{appendix}

\section{Journal of the observations}

\begin{table}[h]
\caption{Summary of the observations. "Rev." is the \inte revolution number.}
\label{tab:observations}
\centering
\begin{tabular}{cccccc}
Rev. & Start & End & Exp. time & Group & State \\
\# & (MJD) & (MJD) & (ks) & & \\
\\
\hline
\\
2678 & 60181.8 & 60182.4 & 51 & 1 & LHS\\
2680 & 60186.6 & 60187.7 & 87 & 2 & LHS\\
2681 & 60188.5 & 60189.7 & 98 & 2 & HIMS\\
2682 & 60192.1 & 60193.0 & 79 & 2 & HIMS\\
2683 & 60194.2 & 60195.2 & 87 & 3 & HIMS\\
2684 & 60197.6 & 60198.3 & 57 & 3 & HIMS\\
2685 & 60200.2 & 60201.0 & 65 & 4 & HIMS\\
2686 & 60203.2 & 60203.7 & 41 & 4 & HIMS\\
2688 & 60207.4 & 60208.2 & 69 & 5 & HIMS\\
2690 & 60213.4 & 60214.3 & 76 & 5 & HIMS\\
2691 & 60214.9 & 60215.7 & 74 & 5 & HIMS\\
2693 & 60221.5 & 60222.1 & 49 & 6 & SIMS\\
2694 & 60222.9 & 60223.5 & 50 & 6 & SIMS\\
2695 & 60226.9 & 60227.6 & 63 & 7 & SIMS\\
2696 & 60229.6 & 60230.3 & 64 & 7 & SIMS\\
2697 & 60230.8 & 60231.5 & 57 & 8 & SIMS\\
2698 & 60233.5 & 60234.3 & 69 & 8 & SIMS\\
2699 & 60236.2 & 60237.1 & 81 & 8 & SIMS\\
\end{tabular}

\end{table}

\section{Compton polarization mode} \label{sec:A-Polar}
The principles for polarization studies with \inte/IBIS have been discussed in several papers \citep[e.g.,][CRB23]{Forot_2007,Laurent_2011,Rodriguez_2015}. In short, we measure the azimuthal angles $\phi$ of each individual scattered photons (corrected for satellite orientation), and group them into $N_p$ bins between $[0,\pi]$, using the $\pi$-symmetry of the probability of diffusion for polarized photons. We create a detector image from each bin of photons according to the position of their first interaction (i.e. their position on ISGRI). We then de-convolve the coded mask from each image to obtain the flux from the source in each polarization bin. This procedure removes the need for background observations and avoids potential contamination from other sources in the field of view, as is always the case for mask deconvolution \citep{Goldwurm_2001}. After careful subtraction of the spurious flux \citep{Forot_2007}, we can fit the resulting polarigram with the following theoretical model:
\begin{equation} \label{eq:1}
N(\phi)=C\ (1+a\cos{(2(\phi-\phi_0))})
,\end{equation}
where $C$ is the mean flux in each bin, $a$ the amplitude of modulation, and $\phi_0$ the privileged diffusion direction. From these parameters we can deduce the PA (measured from north to east on the sky plane) as $\phi_0+\frac{\pi}{2}\ mod\ \pi$ and the PF as $a/a_{100}$, where $a_{100}$ represents the modulation from a fully-polarized source, computed using Monte Carlo simulations\footnote{By convention, all PAs are given here between $-90^\circ$ and $+90^\circ$.}.

With the current PICsIT calibration, we can detect Compton photons with (apparent) energy deposited in PICsIT below the energy threshold of the detector. This is due to long term variations of PICsIT's gain and offset ($E_\mathrm{PICsIT}=\mathrm{gain}\cdot\mathrm{channel} + \mathrm{offset}$) which is not taken into account in the standard software. To correct these effects, we used a fixed gain of $7.05\,\mathrm{keV}\,\mathrm{channel}^{-1}$, based on measurements done with the onboard ${}^{22}\mathrm{Na}$ calibration source (half-life of 2.6\,yrs) during the first years of the mission. We then computed the offset with the 511\,keV internal background line of the satellite itself for each ScW used in our analysis. 

We determine the probability of a detection based on a  probability density function $dP(a,\phi)$ \citep[see, e.g.,][]{Forot_2007}), which we first integrate over $\phi$ while assuming that the source is unpolarized ($a_{src}=0$). We then compute the probability for observations of this unpolarized source yielding a modulation, $a$, which is greater than the modulation we measure, $a_\mathrm{mes}$, by integrating the probability density function from $a_\mathrm{mes}$ to $a_{100}$. The resulting probability is
\begin{equation} \label{eq:2}
p_{\mathrm{unpolarized}} = e^{-k^2\ a_\mathrm{mes}} - e^{-k^2\ a_{100}}
,\end{equation}
where $k^2=N_p c^2/(2\ \sigma_c^2)$, with $N_\mathrm{p}$ the number of angle bins, $c$ the count-rate, and $\sigma_c$ the standard deviation of $c$.

Following CRB23 we consider here that polarization is detected when the S/N of the source in the energy range considered is above $5\sqrt{N_\mathrm{p}}$, and when the unpolarized probability is less than 1\%.
In the cases where we have a good enough signal-to-noise ratio but an insufficient $p_{unpolarized}$, we can reverse (Eq.~\ref{eq:2}) and divide it by $a_{100}$ to estimate the upper limit of $\mathrm{PF}$,
\begin{equation} \label{eq:3}
\mathrm{PF}_\mathrm{up} = \sqrt{\frac{-\ln{(p+\ \exp{(-k^2\ a_{100}^2)})}}{k^2\ a_{100}^2}\ }
.\end{equation}
Here $p$ is the probability that the true $\mathrm{PF}$ is in fact higher than $\mathrm{PF}_\mathrm{up}$. We used $p=1\%$ to match our detection criteria.

\section{Polarization evolution with energy}\label{sec:C-PF}

   \begin{table*}
    \caption{Polarization results in both energy bands. The detection threshold and upper limits are given as explained in Appendix~\ref{sec:A-Polar}.}
    \label{tab:polar}    \centering
    {\renewcommand{\arraystretch}{1.3} 
    \begin{tabular}{cccc|ccc} 
     & \multicolumn{3}{c|}{210 -- 250\,keV} & \multicolumn{3}{c}{250 -- 300\,keV} \\
     \hline
   Group \#  &  S/N & PA   &  PF   & S/N &  PA  & PF  \\
     &     &    &  \%    &  &   & \%  \\
    \hline
1 & 63.7  &  - & < 19  & 29.7  & -  & < 25  \\ 
2 & 102.4 & -45.1 $\pm$ 2.9$^\circ$ & 24 $\pm$ 2  & 55.3 & -51.2 $\pm$ 3.2$^\circ$ & 46 $\pm$ 5  \\ 
3 & 70.5 & -49.2 $\pm$ 5.1$^\circ$ & 42 $\pm$ 8  & 43.5 & -47.4 $\pm$ 6.0$^\circ$ & 50 $\pm$ 10  \\ 
4 & 42.7 & -44.7 $\pm$ 3.9$^\circ$ & 54 $\pm$ 8  & 26.1 & -43.4 $\pm$ 5.8$^\circ$ & 81 $\pm$ 17  \\ 
5 & 39.6 & -30.0 $\pm$ 7.8$^\circ$ & 34 $\pm$ 9  & 20.3  & -  & < 43  \\ 
6 & 15.2 & 3.8 $\pm$ 4.6$^\circ$ & 62 $\pm$ 10  & 9.6  & -  &  - \\ 
7 & 19.5  &  - & < 29  & 10.4  &  - &  - \\ 
8 & 13.7  &  - & < 77  & 7.3  &  - & -  \\

\hline
    \end{tabular}}
\end{table*}

\begin{figure}[h]
    \centering
    \includegraphics[width=\linewidth]{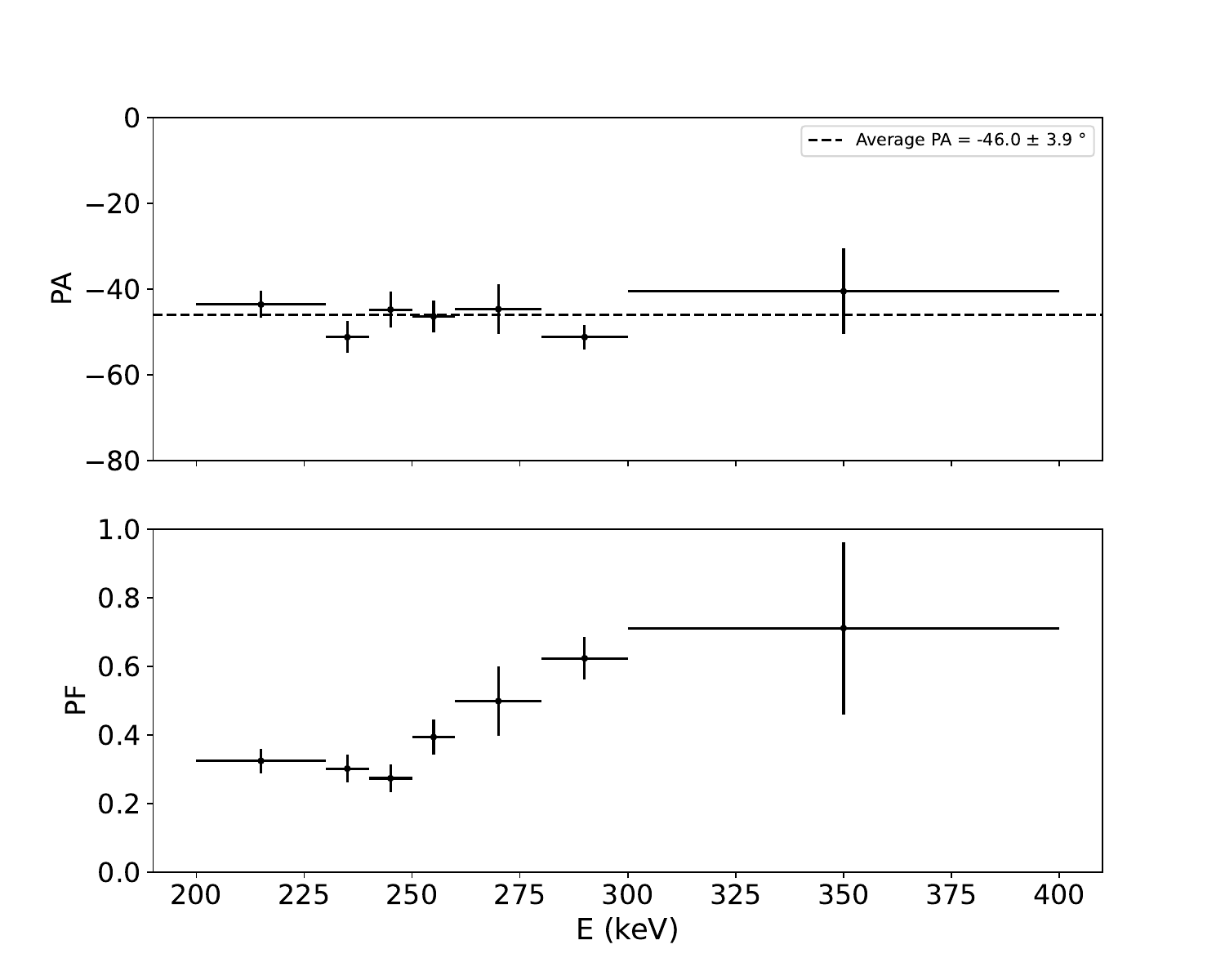}
    \caption{Energy dependence of the J1727 PA and PF in the HIMS.}
\label{fig:S1727_him_pola_spec}
\end{figure}

Despite the variations of PF, we combined Obs.~2 to~4 to look at the energy dependence of the polarization (Fig.~\ref{fig:S1727_him_pola_spec})
The PF increases with energy from 25\% to about 60\% at 300\,keV and stays between 60 and 80\% for higher energies. The PA shows a good stability with energy around a mean value of $-46.0 \pm 3.9^\circ$, suggesting that the polarization in these energy ranges has the same origin.

\end{appendix}

\end{document}